\documentclass[letterpaper,twocolumn,10pt]{article}
\usepackage{usenix-2020-09}
\pdfoutput=1
\usepackage[style=american]{csquotes}
\usepackage{enumitem}
\setlist{noitemsep}

\usepackage{enumitem}
\usepackage{booktabs}
\usepackage{hyperref}
\usepackage{balance}
\usepackage{eurosym}

\def\plaintitle{"It may be a pain in the backside but..." Insights into the impact of GDPR on business after three years}
\def\fancytitle{``It may be a pain in the backside but...''
\\ Insights into the impact of GDPR on business after three years}
\def\plainauthor{Gerard Buckley, Tristan Caulfield \& Ingolf Becker}
\def\plainkeywords{}

\usepackage{hyperref}
\hypersetup{%
 pdftitle={\plaintitle},
 pdfauthor={\plainauthor},
 pdfkeywords={\plainkeywords},
 pdfdisplaydoctitle=true, 
 bookmarksnumbered,
 pdfstartview={FitH},
 breaklinks=true,
 hypertexnames=false
}

\usepackage[style=trad-plain,citestyle=numeric-comp,backend=biber,url=false,
maxcitenames=2]{biblatex}

\newbibmacro{string+doiurlisbn}[1]{%
  \iffieldundef{doi}{%
    \iffieldundef{url}{%
      \iffieldundef{isbn}{%
        \iffieldundef{issn}{%
          #1%
        }{%
          \href{http://books.google.com/books?vid=ISSN\thefield{issn}}{#1}%
        }%
      }{%
        \href{http://books.google.com/books?vid=ISBN\thefield{isbn}}{#1}%
      }%
    }{%
      \href{\thefield{url}}{#1}%
    }%
  }{%
    \href{https://dx.doi.org/\thefield{doi}}{#1}%
  }%
}
\DeclareFieldFormat{title}{\usebibmacro{string+doiurlisbn}{#1}}
\DeclareFieldFormat[book]{title}{\textit{\usebibmacro{string+doiurlisbn}{#1}}}

\AtEveryBibitem{%
        \clearfield{day}%
        \clearfield{month}%
        \clearfield{endday}%
        \clearfield{endmonth}%
        \clearfield{labelmonth}%
        \clearfield{pagetotal}%
        \clearlist{language}%

}

\renewbibmacro*{journal}{%
  \iffieldundef{shortjournal}
    {%
      \iffieldundef{journaltitle}
        {}
        {%
          \printtext[journaltitle]
            {%
              \printfield[titlecase]{journaltitle}%
              \setunit{\subtitlepunct}%
              \printfield[titlecase]{journalsubtitle}%
             }%
         }%
    }
    {\printtext[journaltitle]{\printfield[titlecase]{shortjournal}}}%
}

\bibliography{Year1clean-Ingolf.bib}

\DeclareQuoteStyle[british]{english}
  {\itshape\textquoteleft}
  {\textquoteright}
  [0.05em]
  {\textquotedblleft}
  {\textquotedblright}

\DeclareQuoteStyle[american]{english}
  {\itshape\textquotedblleft}
  {\textquotedblright}
  [0.05em]
  {\textquoteleft}
  {\textquoteright}

\SetBlockEnvironment{quotation} 

\renewcommand{\mkbegdispquote}[2]{\openautoquote}

\begin{document}

\date{}

\title{\Large \bf \fancytitle{}}

\author{
{\rm Gerard Buckley, Tristan Caulfield \& Ingolf Becker}\\
University College London \\
\texttt{\{gerard.buckley.18, t.caulfield, i.becker\}@ucl.ac.uk}
} 

\maketitle

\begin{abstract} 

The General Data Protection Regulation (GDPR) came into effect in May 2018 and is designed to safeguard EU citizens' data privacy.
The benefits of the regulation to consumers' rights and to regulators' powers are well known.
The benefits to regulated businesses are less obvious and under-researched. 

The aim of this study is to investigate if GDPR is all pain and no gain for business. 
Using semi-structured interviews, we survey 14 C-level executives responsible for business, finance, marketing, legal and technology drawn from six small, medium and large companies in the UK and Ireland.  

We find the threat of fines has focused the corporate mind and made business more privacy aware.
Organisationally, it has created new power bases within companies to advocate GDPR.
It has forced companies, in varying degrees, to modernise their platforms and indirectly benefited them with better risk management processes, information security infrastructure and up to date customer databases.
Compliance, for some, is used as a reputational signal of trustworthiness. 

We find many implementation challenges remain.
New business development and intra-company communication is more constrained.
Regulation has increased costs and internal bureaucracy.
Grey areas remain due to a lack of case law.
Disgruntled customers and ex-employees weaponise Subject Access Requests (SAR) as a tool of retaliation.
Small businesses see GDPR as overkill and overwhelming. 

We conclude GDPR may be regarded as a pain by business but it has made it more careful with data. 

We recommend the EU consider tailoring a version of the regulation that is better suited to SMEs and modifying the messaging to be more positive whilst still exploiting news of fines to reinforce corporate data discipline.
\end{abstract}

\section{Introduction}
\label{sec:introduction}

Regulation has long suffered an image problem with the general public for being boring, bureaucratic and unnecessary.
And while this can be true, regulation is a vital lever of government to achieve policy objectives.
Governments regulate business to deliver better outcomes for the economy, the environment and society: for example to correct market failures, to protect people and wildlife from pollution and to safeguard citizens' privacy.
The EU's GDPR is a good example of the latter.

Our focus is on the regulation of data privacy from the perspective of an overlooked stakeholder i.e. business.
Most attention has concentrated on the benefits of GDPR to the regulator in terms of stronger powers and to the consumer in terms of stronger privacy rights.
However little or no research has tracked the benefits of GDPR to business who, after all, have to pay to operate it.

Academic literature, prior to the introduction of GDPR in May 2018, proposed a variety of potential GDPR benefits to business including better data management and analytics, brand enhancement and access to a level playing field.
Since then, however, interest has waned.
Most discussion to date on the benefits of GDPR to business lacks empirical data.
Even recent studies rely on earlier papers and still speak in terms of potential or possible opportunities and benefits.

For this reason, we choose to employ a semi-structured interview technique to ask 14 senior employees in a range of companies what has been the actual impact of GDPR as experienced by them and their companies since 2018. We investigate:

\begin{description}
    \item [RQ1:] What are the benefits of GDPR to business?
    \item [RQ2:] Do the benefits of GDPR fall differently within a business?
\end{description}

After discussing the background to GDPR in Section~\ref{sec:background} and related literature in Section~\ref{sec:litReview}, we give an overview of our methodology in Section~\ref{sec:methodology}. We analyse and discuss the findings in Sections~\ref{sec:findingsAnalysis}~\&~\ref{sec:discussion}. We show there are both direct and indirect benefits accruing to business from GDPR whilst not minimising the ongoing implementation issues. 

We believe this is the first study to analyse the lived experience of GDPR by business since its introduction over three years ago and the first to identify how GDPR has changed the balance of power and decision making within organisations.

\section{Background}
\label{sec:background}

This section provides a quick recapitulation of what is meant by regulation, what are the fundamental principles behind formulating good regulation and how privacy and data protection regulation has developed over time.
It summarises the GDPR, its objectives and the results of subsequent surveys by the EU on the success of its implementation.
Since 2018, we note the dearth of assessment of the benefits to business of GDPR by the EU.
We note a similar lack of follow-up by the professional advisory firms who were active commentators in 2017 \& 18.
Academic assessments follow in Section~\ref{sec:litReview}.

\subsection{What is Regulation?}
\label{sec:whatIsRegulation}

Regulation refers to the mechanisms by which governments set requirements on businesses, and the term \enquote{regulator} refers to a person or authority who develops or administers regulation.
As defined by the Organisation for Economic Co-operation and Development~\cite{oecd_better_2019a}, regulation includes all laws, formal and informal orders, subordinate rules, administrative formalities and rules issued by nongovernmental or self-regulatory bodies to whom governments have delegated regulatory powers.

Governments regulate business to guarantee minimum standards and protections.
Left unchecked, the profit motive of business can lead to damaging behaviours that are detrimental to society e.g., price-fixing cartels, unsafe working conditions, abuses of consumers rights.
While governments may also regulate the actions of individuals, public-sector or civil society organisations, our focus is on the regulation of business and data.

\subsection{The Foundations of Good Regulations}
\label{sec:foundationsOfGoodRegulation}

Regulation brings both benefits and costs.
It can stimulate ideas and can block their implementation.
It can increase or reduce the risk of investing in new products and business models.
It can determine how much funding is available for innovation and how much goes into tick-box compliance.
It can influence consumer confidence and demand and determine whether firms enter or exit a market.

For this reason, most developed economies have policies, procedures and institutions to govern how regulations are developed, administered and reviewed.
While approaches vary, such policies typically affirm the importance of openness, proportionality and fairness~\cite{almond_agile_2020}.

Openness demands transparency and participation in the policy design to ensure regulation serves the public interest and engages all the stakeholders that it affects or who hold an interest in it.
Proportionality demands that the costs of compliance are commensurate with the benefits the regulation is intended to deliver.
Fairness demands that regulatory decisions should be made on an objective, impartial and consistent basis, without conflict of interest, bias or improper influence.
The theory is that this enables businesses to compete on a level playing field, and helps ensure that the best ideas, products and business models are those that succeed.

\subsection{The Evolution of Data Regulation}
\label{sec:evolutionOfDataRegulation}

The right to privacy is part of the 1950 European Convention on Human Rights, which states, \enquote{Everyone has the right to respect for his private and family life, his home and his correspondence}~\cite{europeancourtofhumanrights_european_2021}. As the free flow of information grew widespread, the European Union enacted the 1995 Directive on Data Protection, which imposed minimum standards of personal data protection upon member states and protected the rights of individuals regarding the movement of personal data between EU member states.
However, the law was implemented differently in each EU state, leading to uneven laws and oversight.
In 2011, a Google user sued the company for scanning her emails~\cite{computerweekly_us_2011}.
Soon after, Europe's data protection authority declared the EU needed to update the 1995 directive~\cite{europeancommission_commission_2012}.

\subsection{GDPR Objectives}
\label{sec:gdprObjectives}

GDPR came into force on 25 May 2018~\cite{europeanparliamentandofthecouncil_regulation_2016}, after which all organisations were required to be compliant.
The UK GDPR, post-Brexit, was ruled as adequate by the EU in June 2021.

Unlike its predecessor, GDPR is an EU Regulation and not a Directive.
This means it has binding force in every member state and there is no discretion over how it is transposed into national law.

The primary purpose of GDPR is to define standardised data protection laws for all member countries across the European Union. It was intended to:

\begin{itemize}
 \item 	Increase privacy and extend data rights for EU residents.
 \item 	Help EU residents understand personal data use.
 \item 	Address the export of personal data outside of the EU.
 \item 	Give regulatory authorities greater powers to take action against organisations that breach the new data protection regulations.
 \item 	Simplify the regulatory environment for international business by unifying data protection regulations within the European Union (a.k.a. the level playing field (LPF)).
 \item 	Require every new business process that uses personal data to abide by the GDPR data protection regulations and Privacy by Design rule.
\end{itemize}

It has strict rules such as the rights for data subjects to access their own data (known as SARs), to be forgotten and to expect affirmative consent.
It applies to companies inside and outside the EU if they hold personal data belonging to EU citizens.
And it has tight data breach notification requirements and hefty fines of up to four percent of an organisation's total worldwide annual turnover if found in violation.

GDPR is strong on the obligations of business. It makes no reference to any benefits to business.

\subsection{GDPR Scorecard}
\label{sec:gdprScorecard}

The EU has commissioned a number of surveys since the GDPR was applied.
We highlight three surveys here: the 2019 Eurobarometer, the 2019 SME Survey and the 2020 EU Self-Evaluation Report.

The 2019 EU Barometer 487a~\cite{europeancommission_general_2019} found that:

\begin{itemize}
\item Over 66\% of EU citizens have heard of GDPR, over 50\% have heard of their rights under GDPR, and almost 60\% have heard of their data regulator.
\item A majority feel they have partial control over the information they provide online. Only 20\% say they see the T\&C's to the collection and use of their personal data online, and only 13\% say they read privacy statements in full.
\item 56\% say they have tried to change their default privacy settings. The most common reason for not doing so are that users trust the pre-set privacy settings (29\%) or that they do not know how to do it.
\end{itemize}

The 2019 GDPR Small Business Survey was run by Proton Technologies AG~\cite{gdpr.eu_gdpr_2019}. Part-funded by a EU Horizon Project, it found:

\begin{itemize}
\item Millions of small businesses still do not comply with the GDPR.
\item Encryption technology is still not widely understood.
\item Small businesses want to comply and have invested heavily on GDPR compliance. 
\end{itemize}

On June 24, 2020, the European Commission (EC) submitted its first report on the evaluation and review of the GDPR to the European Parliament (EP) and Council~\cite{commission_communication_2020}.
The report is required under Article 97 of the GDPR and will be produced at four-year intervals going forward. In its report, the Commission concludes that generally the GDPR has successfully met its objectives, namely those of strengthening personal data protection and guaranteeing the free flow of personal data within the EU. It identified a number of areas for improvement, including:
\begin{itemize}
\item	Fragmentation between member states: differential interpretation of discretionary details
\item	Uneven enforcement: different \enquote{data protection cultures}, different budgets \& resources
\item	Unforeseen Issues with Emerging Technologies: AI, IoT or facial recognition
\item	Unused Potential of Data Portability Rights: to avoid unfair practices and lock-in effects
\item	Adequacy Decisions: Pending third country regimes such as South Korea and UK
\item	Extra-territorial Reach: \enquote{This approach should be pursued more vigorously in order to send a clear message that the lack of an establishment in the EU does not relieve foreign operators of their responsibilities under the GDPR.}
\end{itemize}
Whilst this report is akin to the EC marking its own homework and not an impartial external assessment, it is still a useful checklist of where the EC sees shortcomings in GDPR.

\subsection{Gap in GDPR Scorecard}
\label{sec:gapInScorecard}

Our search has revealed a significant gap in the assessment of GDPR.
There seems to be no equivalent to the EC's four-yearly evaluation and review from the perspective of one important stakeholder: the regulated businesses that handle customer data. 

In the run-up to GDPR going live in 2018, there was a flood of surveys, studies and benchmarking reports by IT vendors and professional services firms. Since then, they have dried up. 

One exception is the EU Multistakeholder Expert Group.
Set up in 2017, it assists with identifying the potential challenges in the application of the GDPR from the perspective of different stakeholders, and to contribute to the EC's evaluation of GDPR in 2020.
It is composed of up to 27 members drawn from trade and business associations, NGO's, academics, legal practitioners and privacy advocates. It is quite technocratic.
To give a flavour, their \textquote{contribution addressed topics such as the impact of the GDPR on data subjects' rights, the conditions for a valid consent under Article 7(4) of the GDPR, the one-stop-shop mechanism, the principle of accountability and the risk-based approach, data protection officers' ('DPOs'), the relationship between controllers and processors, and the development of Standard Contractual Clauses for the transfer of personal data}~\cite{europeancommissiondirectorategeneralforjusticeandconsumers_eu_2020}. Benefits analysis is not part of its mandate.

One of the few non-EU follow up surveys was a survey by Deloitte's \textquote{A new era for privacy: GDPR six months on}~\cite{deloitte_gdpr_2018}. The headline was that consumer awareness has risen and 48\% of organisations had made ``significant'' investment to improve their compliance. In addition:
\begin{itemize}
\item	70\% of organisations had increased staff focused on GDPR compliance. 
\item	GDPR had global reach: 22\% of organisations from EU countries maintained the same number of staff, compared to 20\% from non EU countries. 
\item	92\% of organisations claimed confidence in their ability to comply with GDPR in the long term. 65\% of organisations felt they had enough resources to comply. 
\item	78\% had invested in new data loss prevention and 71\% in unstructured data scanning.
\end{itemize}

When the EU first introduced GDPR, it was quite blunt about its objective - namely, the protection of the consumer.
On the other hand, professional services firms, technology providers and academia talked up the benefits of GDPR to business. Since 2018, roles have reversed.
The EU sprinkles its GDPR surveys with positive messages about the benefits of a level playing field and GDPR compliance whilst the service industry has fallen silent.

\section{Literature Review}
\label{sec:litReview}

There is no shortage of GDPR studies. A Google Scholar search of General Data Protection Regulation will yield circa 3 million hits.
Limit the search to papers published after GDPR went live in 2018 however and interest drops precipitously.
Search for papers that contain the two keywords ``GDPR'' and either ``success'' or ``benefits'' anywhere in the text yields less again. As we narrowed the search, we quickly reached zero hits for keyword combinations such as ``GDPR business benefits'' or ``GDPR consumer benefits'' or even ``benefits of GDPR to business''. The lack of curiosity about GDPR's benefits to business after 2018 is curious. 

In this section, we review the plentiful literature on the implementation challenges of GDPR. We then examine papers that contained the word pair ``GDPR success'' and ``GDPR benefits'' anywhere in their text as well any relevant papers from multiple rounds of backward and forward searches.
Most of these do not talk to our topic because they are not interested in how GDPR might deliver value or return from a business perspective. Two papers that are aligned with our research (including one systematic literature review) still refer back to pre-GDPR research.
We conclude that given the newness of GDPR, there are still few scientific follow-up studies. 

\subsection{GDPR Implementation Challenges}
\label{sec:gdprImplementationChallenges}

Unlike benefits, there is a surfeit of studies on the challenges of GDPR.
It is a complex regulation~\cite{freitas_gdpr_2018}, it fails to specify technical solutions~\cite{tikkinen-piri_eu_2018} and it involves subjectivity~\cite{agarwal_dealing_2016}. Compliance can be expensive~\cite{tikkinen-piri_eu_2018,addis_general_2018}.
Companies may need extra administration staff and expert DPO staff~\cite{lindgren_gdpr_2016}, extra employee training and face difficulty recruiting and retaining these people~\cite{khan_need_2018}.
Regulatory restrictions may impact an organisations performance~\cite{vandermarel_methodology_2016} and persuade some to cut back their service offering in the EU to avoid it~\cite{allen_economic_2019}. 

GDPR brings increased technical complexity~\cite{bennett_european_2018,dubrova_challenges_2018,politou_forgetting_2018}.
Data portability~\cite{kaushik_data_2018} as well data consent, rectification and deletion processes will require technical and organisational investment~\cite{dubrova_challenges_2018}. Data erasure (aka the right to be forgotten) is seen as particularly problematic for bigger companies~\cite{dehert_right_2018,dove_eu_2018}. System and process audits~\cite{dubrova_challenges_2018} and recruiting more cybersecurity professionals will require more investment.
Clamping down on how personal data is handled may slow down the development and application of emerging technologies such as IoT and blockchain~\cite{li_impact_2019,wallace_impact_2018}. 

\subsection{GDPR Success Studies}
\label{sec:gdprSuccess}

Unlike challenges, there is a dearth of research on success. 
Under ``GDPR success'', the most relevant literature has a regulator or regulatory success focus rather than any reference to business success. 
Thus, Oxford Analytica's appraisal of GDPR on its first anniversary~\cite{oxfordanalytica_europe_2019} looked at key shortcomings such as ensuring the compliance of business beyond ``big tech'', concern that public awareness of the GDPR in smaller EU states will lag that in larger states and criticism of the Irish regulator if it failed to demonstrate a clearer commitment towards robust regulation. 
Sanders, in \textit{The GDPR One Year Later}~\cite{sanders_gdpr_2018} suggests the key to the GDPR's success requires data protection officials and judges to seriously evaluate situations in which privacy and freedom of the press appear to conflict. 
Kessler in \textit{Data Protection in the Wake of the GDPR: California's Solution for Protecting ``the World's Most Valuable Resource''}~\cite{kessler_data_2019} argues that the United States should adopt a federal standard that offers consumers similarly strong protections as the GDPR. 

\subsection{GDPR Benefits Studies} 
\label{sec:gdprBenefits}

Under ``business benefits'', the literature has a more business focus albeit about the dis-benefits of GDPR. \textit{The Economic Impact of the European Reform of Data Protection} is a 2015 paper by M Ciriani~\cite{ciriani_economic_2015} of the Regulatory Office of the giant French mobile phone operator Orange. She argued that the extraterritorial application of European law would promote a level playing field within the European market.
However, with the exception of the GDPR's impact assessment conducted by the European Commission, she claimed the literature she had examined shows that the costs of GDPR's adoption might offset the efficiency gains.
She expressed concern that increasing the administrative burden might not help improve the competitiveness of European digital service providers, such as her employer.
Flexible ex-post effects-based accountability would help industry. 

Sarah Shyy, in the self-explanatory \textit{The GDPR's Lose-Lose Dilemma: Minimal Benefits to Data Privacy \& Significant Burdens on Business}~\cite{shyy_gdpr_2020} argues that GDPR fails to promote consumer privacy because, in today's data collection practices, consumers are forced to accept an online company's privacy policy and data collection practices.
Meanwhile, the GDPR has disadvantaged SMEs by imposing cost-prohibitive measures, hindering SMEs growth, and spurring SMEs to exit the market.
Rather than copying the GDPR model, she argues US lawmakers should learn from the GDPR's failings and adopt regulation that is both more effective in protecting consumer privacy and less burdensome on businesses. 

\subsection{GDPR Benefits to Business Studies}  
\label{sec:gdprBenefitsBusiness}

We identify two papers published in 2019 that look at GDPR from a business perspective. We find neither are based on academic research conducted after 2018. 

\citeauthor{poritskiy_benefits_2019} asks what are the main benefits offered by GDPR for IT companies~\cite{poritskiy_benefits_2019}.
Their study used a questionnaire to seek feedback on eight benefits (and nine challenges) identified by other researchers.
They concluded the two most significant benefits were trust (consumer confidence) and legal clarification.
These two benefits confirm the view of two of their sources, Bilyk~\cite{dubrova_challenges_2018} and White~\cite{whites_general_2018}.
The Bilyk study references a blog on theappsolutions.com website credited to a different author and the White study appears to reference a (now) broken link to a GDPR news report.
Another two of the eight benefits of GDPR, better decision-making and better risk-assessment, are also credited to Bilyk.
Two more of the eight benefits, increased security of products/services and increased quality of documentation, are credited to Krikke et al.~\cite{krikke_gdpr_2019} which appear to reference brochure-style content on the site of the law firm Stibbe.com.
Another benefit, create new competitive advantages, is credited to Dellie~\cite{dellie_gdpr_2019} which references a blog on the ITASCA.org site.
Two further benefits, minimisation of the collected personal data and improved data management processes, are credited to Fimin~\cite{fimin_council_2018} which references an article in Forbes magazine by the CEO of cybersecurity firm.
The eight benefits surveyed in the questionnaire may indeed be real but the cited evidence behind them is not based on any qualitative or quantitative data. 

In the second paper, Teixeira et al.~\cite{almeidateixeira_critical_2019} conducted a systematic literature review to identify the critical success factors that contribute to the implementation of GDPR.
One of the research questions was \enquote{What are the benefits of complying with GDPR?} 

Their review identified four potential areas of benefit:
\begin{itemize}
\item proper data management 
\item use of data analytics 
\item cost reduction
\item increase in reputation and competitiveness. 
\end{itemize}
Regarding data management, \citeauthor{lopes_implementation_2018} view GDPR as an opportunity for companies to document their processes and procedures~\cite{lopes_implementation_2018}. \citeauthor{presthus_gdpr_2018} sees it as an opportunity to cleanse and audit personal data to cap any liability to abuse of personal data~\cite{presthus_gdpr_2018} and similarly, \citeauthor{skendzic_general_2018} view GDPR as an opportunity to bring data consistency across the organisation~\cite{skendzic_general_2018}.

Another benefit of better data management is better data analytics.
\citeauthor{garber_gdpr_2018} argue data-driven insights will help inform companies optimise their business processes and identify new business development opportunities~\cite{garber_gdpr_2018}. Another benefit of better data management is lower costs.
Others argue enhanced data management will eliminate surplus data, redundant data and, thus, data storage costs~\cite{miglicco_gdpr_2018,beckett_gdpr_2017,perry_gdpr_2019}.
\citeauthor{obrien_privacy_2016} reports it could reduce costs by up to 2.3B EUR per annum according to estimates by the European Commission~\cite{obrien_privacy_2016}.
\citeauthor{beckett_gdpr_2017} postulates that GDPR compliance and safe data governance skills may enhance a company's trustworthiness and generate new business and new customers~\cite{beckett_gdpr_2017}.
\citeauthor{tikkinen-piri_eu_2018} argue the adoption of GDPR may give a competitive advantage to organisations~\cite{tikkinen-piri_eu_2018}.
\citeauthor{garber_gdpr_2018}  and \citeauthor{miglicco_gdpr_2018} believe compliance may also boost an organisations' performance by improving operational efficiency~\cite{garber_gdpr_2018,miglicco_gdpr_2018}. 

It is noteworthy that all the papers in the review, bar one~\cite{perry_gdpr_2019}, are dated 2016 to 2018 and couched in terms of potential benefits that may help organisations to realise an ideal state.
The only article from 2019 is credited to R Perry, VP of Product Marketing at ASG Technologies which is a provider of enterprise information management solutions~\cite{perry_gdpr_2019}.
There is no post-2018 data to support the projected benefits. 

\subsection{Motivation}
\label{sec:motivation}

Given GDPR is a relatively recent regulation, there are still few scientific follow-up studies. Neither the 2019 systematic literature review nor our review here were able to identify or substantiate specific benefits to business of implementing GDPR. Hence our research aim is to obtain information on the actual impacts of GDPR and how those effects are felt across the organisation.  

\section{Methodology}
\label{sec:methodology}

This study asks what are the benefits of GDPR to business and if the benefits fall differently across the organisation.
Due to the newness of the issue and the absence of reliable data, it takes a qualitative method, thematic analysis, to identify common topics or ideas that come up repeatedly in interviews.

\subsection{Data Collection and Analysis}
\label{sec:dataCollection}

Data was collected via a series of semi-structured one-to-one interviews with business executives who work in companies that handle customer data.
The agenda was developed by drawing on the literature review and capturing the predicted benefits and challenges. These were used as a checklist to ensure all the talking points were covered if they did not come up unprompted in response to the initial open questions.
The interviews were recorded, transcribed and coded for themes that arose in the course of the project.
The aim was to cover a range of small, medium, and large companies and a range of roles and departments (IT, Legal, Marketing, General Management) to maximise the sample's representativeness. 

The questions were continuously reflected upon and adjusted during the project. The research team reflected on the approach after each interview.
The analysis used an inductive and iterative approach to uncover and refine the themes in the responses, informed by the benefits and dis-benefits identified in the literature review.
One author conducted all the interviews to maintain consistency. 

\subsection{Interviews} 
\label{sec:interviews}

Each interview had four parts: 
\begin{enumerate}
 \item Open contextual questions about the individual, their job title, their company, its size and sector.
 \item Open questions about how GDPR had affected their job, department and company. 
 \item Targeted questions that explored a check-list of suggested advantages of GDPR.
 \item Targeted questions that explored a check-list of suggested disadvantages of GDPR.
\end{enumerate}
The interview guide can be found in Appendix A.

\subsection{Sample Characteristics}
It is difficult to recruit business people to dedicate time to an interview at the best of times. It is doubly difficult if the target is senior management and if the topic is a commercially sensitive matter. Based on the researcher's networking skills, 14 senior executives agreed to be interviewed across six companies based in the UK and Ireland. The breakdown is shown in Table~\ref{tab:organisations}.

\begin{table}[htb!]
\centering
\begin{tabular}{llll}
\toprule
\textbf{Size} & \textbf{Sector} & \textbf{Job Title}                                        & Labels                                               \\ \midrule
Small         & Technology      & MD                                                        & P1                                               \\
              &                 &                                                           &                                                     \\
Small/Med & Publishing & \begin{tabular}[c]{@{}l@{}}CEO\\ MD\\ FD\end{tabular}           & \begin{tabular}[c]{@{}l@{}}P2\\ P3\\ P4\end{tabular}          \\
              &                 &                                                           &                                                     \\
Medium        & Manufacture     & \begin{tabular}[c]{@{}l@{}}CEO \\ IT MD\end{tabular}      & \begin{tabular}[c]{@{}l@{}}P5\\ P6\end{tabular}  \\
              &                 &                                                           &                                                     \\
Large         & Drinks          & \begin{tabular}[c]{@{}l@{}}Legal\\ Marketing\end{tabular} & \begin{tabular}[c]{@{}l@{}}P7\\ P8\end{tabular}  \\
              &                 &                                                           &                                                     \\
Large         & Law             & \begin{tabular}[c]{@{}l@{}}Legal\\ CIO\end{tabular}       & \begin{tabular}[c]{@{}l@{}}P9\\ P10\end{tabular} \\
              &                 &                                                           &                                                     \\
Large     & Bank       & \begin{tabular}[c]{@{}l@{}}Legal\\ CISO\\ CMO\\ PR\end{tabular} & \begin{tabular}[c]{@{}l@{}}P11\\ P12\\ P13\\ P14\end{tabular} \\
\bottomrule
\end{tabular} 
\caption{The organisations and interviewees labelled P1--P14}
\label{tab:organisations}
\end{table}

\subsection{Ethical Considerations}
\label{sec:ethicalConsiderations}

The authors' departmental Research Ethics Committee approved this study. It is designed to include pseudonymity, confidentiality and informed consent. The study does not identify individual participants. All identifiable information was stripped from the transcripts and the recordings were subsequently deleted. Some quotes were altered or redacted to mask details. The participants were aware of the research's purpose, the researchers involved, and their role in it. Participants were not offered any compensation for participating.

\section{Findings \& Analysis}
\label{sec:findingsAnalysis}

We are interested in what the benefits of GDPR to business are and how the benefits, if any, fall differently within a business.

In this section, an expansive definition of `benefit' is taken because benefits in business come in many guises.
Typically, companies will classify benefits as either direct or indirect.
Direct benefits have a clear cause and effect relationship, whilst indirect benefits are less clear cut.
A direct benefit will generate new revenue or reduce costs and is quantifiable.
An indirect benefit, sometimes called a soft benefit, may be less tangible and defy direct measurement. Benefits may be planned or unanticipated.
Benefits can also be in the eye of the beholder: what advantages one part of an organisation can disadvantage another.

Our discussion is based on themes that were generated by analysing and abstracting the interviews.
Four direct benefits are identified:
\begin{itemize}
   \item 	Cultural: Privacy-Aware Mindset,
   \item 	Technical: Spur to Change,
   \item 	Reputation Signal,
   \item 	Standardisation.
\end{itemize}

Two indirect benefits were identified: 
\begin{itemize}
    \item Powerful in-house GDPR advocacy, 
    \item Improved data management and information security. 
\end{itemize}

While the primary focus of the research was benefits, we identify many outstanding challenges with implementing GDPR:
\begin{itemize}
    \item 	New Business Development is Harder,
    \item 	Direct \& Indirect Costs,
    \item 	Grey Areas of Law,
    \item 	The Data Audit Dividing Line,
    \item 	The Weaponisation of SARs.
\end{itemize}

The participants were invited to suggest how GDPR could be made better. We review their feedback in the final section.

\subsection{Direct Benefits of GDPR to Business}
\label{sec:directBenefits}

\subsubsection{Privacy-Aware Mindset}
\label{sec:directPrivacyAwareMindset}

All interviewees stressed the importance of protecting data and privacy. It came across as contrived. On closer questioning, the sincerity and the motivation became clear. It is driven by fear. GDPR fines focus the mind. For severe violations listed in Art. (5) GDPR, the fine framework can be up to \euro20 million or up to 4\% of global turnover, whichever is the higher. As one executive observed \textquote{It does put the whole of the organisation into a different mindset} (P12).

The existential threat is felt by large and small players alike. A senior executive in one of the larger companies said, 
\begin{displayquote}[P13]
We do simulated exercises around crises \textelp{} and they always come down to a cyber hack and data leakage, and data is what we run our business by. So generically, it is the thing that keeps me awake at night the most, and it is the one thing that could blow our company up.
\end{displayquote}
The Finance Director of an SME put the effect of a fine more pointedly, \textquote{we're running on fumes most of the time anyway, so any little thing could push us over the edge financially} (P6).

GDPR has made executives more aware of data privacy both at a corporate and at a personal level.
\begin{displayquote}[P4]
We were very aware of its arrival \textelp{} it approached us as something of a tidal wave of regulation because we knew that the sanctions against companies that failed to adhere were going to be quite stiff. And we also understood that it was important. We all have personal lives and know what it's like when we have interference and intervention and unnecessary and unsolicited approaches by organisations. 
\end{displayquote}
Put another way, treat others' data as you would your own.

GDPR has changed companies' data use behaviour. A marketer put GDPR's impact as follows: 
\begin{displayquote}[P13]
{As painful as it might be, fundamentally what it allows us to do is to understand our customers desired level of engagement with our company. \textelp{In the} wild west before GDPR came along \textelp{} you didn't have to bother about things like marketing permissions and things like that. \textelp{You used to} have a mass of customers; you used to contact them through whatever means whenever you wanted to, and some were more receptive to that than others}.
\end{displayquote}
In other words, as a business practice, no more spam.

GDPR has changed the corporate mindset. It has raised \textquote{awareness within the business around personal data, the importance of protecting it and treating it in specific ways}. (P7) It \textquote{has led to a better understanding of why we hold data} (P7). It stops people from hoarding data and gets rid of the \textquote{just in case mentality} (P9).

It has also \textquote{raised awareness within the business from a cyber perspective \textelp{} which has resulted in us procuring cyber insurance} (P9). \textquote{Security is tighter now \textelp{} in terms of encryption \textelp{} we've tightened down access} (P7).

In sum, GDPR has made companies become more responsible. As one CEO put it, \textquote{I suppose we just are a little bit more careful what we use the information for} (P2).

\subsubsection{Spur to Change}
\label{sec:directSpurToChange}

New regulation like GDPR may require companies to change their people, processes or technology.
It depends on how close their model of operation was already in alignment with the new regulation. Suppose a company is forced to buy a new data system to satisfy GDPR and the new system delivers a whole host of new efficiencies and cost-savings. In that case, it is difficult to argue that they are direct business benefits of GDPR. After all, the company could have used that same money for something more value generative such as new product development or expansion into new markets. However, if a company is forced to face up to longstanding issues that it knew had to be fixed or it would degrade or die, and if GDPR is the spur to make that investment finally, then it is arguable that the spin-off benefits that accrue from that investment can be regarded as a direct benefit to business from GDPR.

Out of the six companies in the study, two made no changes to their IT infrastructure. One tweaked their data classifications. One was created after 2018 and was designed from the outset with GDPR in mind, and the remaining two were spurred to make fundamental changes. In both cases, GDPR was responsible for accelerating the necessary change.

One of the SMEs said the most significant benefit of GDPR was \textquote{getting things in order} (P4). \textquote{We had enough spreadsheets to fit in a football field} (P4). They moved everything onto the cloud, went paperless, slashed costs and reduced headcount by 2/3rd. In effect, GDPR meant \textquote{driving the digitalisation and automation of a lot of systems \textelp{} and the restructure of the organisation} (P6).

In contrast, one of the larger companies had already concluded that data-driven marketing \textquote{is the way of the future} (P8). It used GDPR as an opportunity to centralise all country customer databases in Global HQ, standardise the data input and output processes, tighten access control and upgrade information security. They required all customers to re-opt in as part of a campaign to be GDPR-ready. They programmed standards into their workflow to enforce GDPR principles such as data minimisation and data retention periods and made it apply worldwide. As an example, the company now has an automated rule that flags and deletes prospect data if they have not been touched or communicated with a prospective customer after a year. It also serves as a feedback loop between country management and GHQ. Why have you neglected to contact these prospects? Did a mass campaign target the wrong market?

Did GDPR spur innovation? Possibly. All six companies initially said no when asked directly and then gave examples that sounded curiously like a new product or service. The technology SME described how their supplier had added a self-service facility so that clients could interrogate and edit their own details, i.e. a do-it-yourself SAR. The IT outsourcing suppliers freely admitted it had increased existing business and created new business as they raced to develop new services to respond to clients' GDPR-related demands. Likewise, the law firm had added extra staff to handle the GDPR workstreams, opened a new branch office in the US to advise local firms on it and expanded a GDPR-compliant legal technology platform service to its clients who needed to pool and overview legal matters internationally. The drinks company created a ``search squad'' to optimise clients' journeys through the GDPR tick boxes of its website. The bank said the number of companies in the GDPR-servicing sector had expanded and granted them a wider selection of compliance systems.

\subsubsection{Reputational Signal}
\label{sec:directReputational}

Reputation management is about sending the right signal to the right stakeholder. Does GDPR compliance by a company, communicated via their public privacy policies and online cookie consent notices, enhance a company's brand and reputation? Do consumers trust it more? Interviewees tied themselves in knots considering this. Many started with a flat \textquote{No} or, slightly less dismissively, \textquote{I don't think it is high up in people's minds \textelp{} since the legislation is no longer a choice and we all have to be compliant} (P2) or \textquote{it's a minimum standard} (P3). The recognition \textquote{it's a necessary element of doing business} (P1) morphed into \textquote{I think failure to do it can impact negatively} (P7) and \textquote{It is a hygiene factor. If you are not GDPR compliant, you've got a problem} (P10).

The concept of hygiene factors dates to psychologist Frederick Herzberg's two-factor theory of worker motivation~\cite{herzberg_motivation_2017}, which marketers later adopted to mean the basic set of values that customers expect to be in place for any business or service they consider purchasing. In mathematics, it would be described as a necessary but not sufficient condition. \textquote{Everyone wants to see that you are obeying looking after your data} (P4). When asked about trust, a lawyer said, \textquote{The customer expectation is higher. I'd say expectation more so than trust is higher} (P7). In contrast, a marketer said, \textquote{People are looking for brand purpose. They're looking for brands with meaning. They're looking for a brand with authenticity. They're looking for brands that do the right thing} (P8). Referring to marketing communication, another said, \textquote{From a client point of view, they know that you are only sending them stuff that they want to receive} (P14).

So, some companies regard GDPR as a box to be ticked and some regards it as a signal and trust builder. It is used to send a signal of \textquote{reassurance} (P14) that the consumer will not be spammed. It is used to say we care about your data, and you can trust us. In fact, one CISO believed that their ISO27001B certification was an indirect benefit to the customer because it tells the customer \textquote{we actually take security seriously} (P12). In an online world where service experience is relatively undifferentiated, reputation is the differentiator and GDPR compliance is now part of it, whether companies chose to exploit it or not. 

\subsubsection{Standardisation}
\label{sec:benefitStandardisation}

Standardisation is often seen as a positive output from regulation. The theory is that technical standards facilitate faster economies of scale on the supply side and provide the comfort of mind to encourage more rapid take-up on the demand side. GDPR might seem an unlikely exemplar, but three cases came to light that delivered direct business benefits. 

In the first case, an SME that did a lot of business with the public sector described how time-consuming it used to be to bid for a new project because each \textquote{organisation would write their own requirements around privacy} (P1). Now GDPR has made responding to formal tenders for new business a quick box-ticking exercise instead. In the second case, our drinks company that exports most of its product overseas also realised time savings. It used to have to consult individual countries on the online and offline product packaging before every new product launch. Now GDPR has made policy wording consistent across the EU. In the third case, our bank felt GDPR had eliminated a competitive disadvantage. As the banker wryly observed, now his competitors had to obey the same laws as he was forced to by his \textquote{strong} in-house risk managers who were \textquote{militant} in their \textquote{oversight} (P13).

These may qualify as examples of the level playing field benefit that the EU has touted as a benefit of GDPR.

\subsection{Indirect Benefits}
\label{sec:indirectBenefits}

\subsubsection{Powerful GDPR Advocates}
\label{sec:indirectPowerfulAdvocates}

Andrew Jackson, seventh president of the United States, is credited with the saying \textquote{money is power}. Within companies, this translates into budget is power, and nowhere is this more apparent than the power that GDPR has conferred on specific roles within companies to invest in compliance. One in-house counsel was quite frank: \textquote{I am a boring lawyer, but I think the fact there's robust legal obligations has made business ensure compliance at a speedier rate than usual. \textelp{} The level of fines makes for a great headline when you're running training and trying to get everyone's appreciation} (P7). Another enforcer put it succinctly: \textquote{Data breaches \textelp{} gives everyone an incentive to listen \textelp{} the 4\% \textelp{} is hanging over the heads of the board} (P11).

GDPR has transformed the authority of the department responsible for it--–usually Legal, Risk or Compliance–--and made it an essential player in corporate data-related decision making: \textquote{It has raised awareness of the compliance team. \textelp{} People take compliance a lot more seriously} (P9).

It might seem the main benefit of a beefed-up GDPR-legal resource is fine limitation. Still, the lawyers would also point to reduced paper storage costs, greater awareness of cyber risk leading to the taking out of cyber insurance, tighter scrutiny of third-party supplier contracts and more attention to where the data in the cloud resides to ensure the EU GDPR regime covers it.

\subsubsection{Generic Benefits of IT Implementation}
\label{sec:indirectGenericIT}

The other budgetary beneficiary is the IT department. One CISO (P12) believed GDPR \textquote{did raise the bar for visibility of information security \textelp{} in the past \textelp{} it was regarded as a nice to have. \textelp{} Not many companies actually had a security department}. This CISO also thought that \textquote{GDPR focused people's minds that if you let the data get out, then it could conceivable bring down the company}. Another CISO (P10) described how they work with risk and compliance to document risk and list controls they had against those risks so that when they suffer a breach, an inevitability in their view, they can demonstrate to the regulator they had made a proportionate investment to meet their obligations to the spirit and letter of the legislation.

Thus, the budget has been invested in information security infrastructure, resilience, and eliminating single points of failure. Another focus is security awareness training for the workforce. The investment has resulted in streamlined processes, efficiencies and cost savings. It has motivated companies to take a holistic view of security rather than \textquote{sticking the firewall in the way} (P12). It has meant that the customer database is constantly cleansed and deduplicated to ensure client notification preferences are up to date, which in turn means the advertising is targeted at customer and prospective customers who are genuinely interested in the company's product or service. As one marketer put it, prior to cleaning up our data and duplicates, \textquote{we used to have multiple versions of the truth} (P13).

\subsection{Challenges}
\label{sec:challenges}

\subsubsection{New Business Development is Harder}
\label{sec:challengeBusinessDevelopment}

A key part of new business development is identifying, qualifying and converting suspects into prospects and prospects into customers. A pipeline of leads is generated via a variety of means such as advertising, social media and email marketing campaigns.

The biggest drawback of GDPR for one SME was \textquote{finding effective ways to find new customers} (P1). He recounted wistfully how, before GDPR, their resellers made it a precondition that users had to agree to receive spam before their service was activated. Even though it has been against EU law to send unsolicited commercial emails or texts for almost 20 years, it seems to have taken the introduction of GDPR to get the message finally through to business because it changed the rules of consent and strengthened people's privacy rights.

Smaller companies felt GDPR had little effect on them since they were never great spenders on advertising in the first place. However, on exploring the application of the data minimisation principle, there was a dawning realisation by all the SMEs that it had affected them. In practice, they had stopped asking for more information than strictly required so that it could be used again in later campaigns. Previously, they used to periodically re-market to historic enquirers, ex-customers, or lapsed subscribers as a matter of routine. 

Larger companies thought it had made their marketing more targeted and effective because they only communicated with genuinely engaged consumers who had already opted-in to receive marketing communications: \textquote{GDPR forces us to categorise customers according to their wishes and to segment the communication we send them} (P13).

The flip side for marketers is that it made it harder to build the brand if they were only allowed to talk to the \textquote{converted} (P8). It also made it harder for IT in large companies if they had multiple streams of leads (referrals from the parent company or associate companies, web enquires, responses to marketing campaigns, Facebook, Linkedin, Twitter, Google ads) because they had to deduplicate the customer to ensure their preferences were captured correctly and thereby avoid complaints about receiving unwanted marketing communications. 

\subsubsection{Direct \& Indirect Costs}
\label{sec:challengeCosts}

How companies experience the cost of regulation varies widely. One SME remarked he expected the costs to be more, but their only cost was the \textquote{minimal} ICO (UK Regulator) fee (P1). Another SME believed their costs had gone up because they had moved everything to GDPR-compliant cloud providers and assumed their transaction fees included a GDPR component. In general, apart from explicit GDPR-related costs such as cookie notice plug-ins, SMEs found it difficult to pinpoint additional costs.

Larger companies found it easier because they had made more extensive investments in systems, processes and manpower. One company estimated \textquote{15\% of our legal budget in the last year was probably on data protection} (P7). Another put it at less than 5\% (P10). In addition to direct costs, there were indirect opportunity costs. A Global IT Director described GDPR as \textquote{stifling} and \textquote{distracting} (P10). He complained that GDPR projects always trumped other innovative projects such as process automation. Another complained that they had lost business due to GDPR because it made the company so reluctant to share referrals or client information with associate companies in other EU countries.

Attitudes to the added expense vary depending on the department. Marketing sees it \textquote{as an additional burden} (P11). They complain \textquote{they have no time and no budget for it} (P11) and it makes their campaigns uncompetitive against players willing to sail closer to the wind. In contrast, IT see the bureaucracy as \textquote{a cost worth bearing} (P12) if it brings \textquote{sensibility} (P12) to an organisation.

\subsubsection{Grey Areas of Law}
\label{sec:challengeGreyAreaOfLaw}

Unsurprisingly, non-legal and legal interviewees had different perspectives on the state of the law. Most SME management did not have an opinion apart from a shared consciousness that they lacked in-house compliance knowledge. Some expressed worries about loose data hygiene by staff working from home. Some worried about SARs and how much disclosure was required. All thought they had outsourced responsibility for security under GDPR compliance to their GDPR outsourcers. 

Participants who did have contact with GDPR complained simultaneously that GDPR was over-prescriptive and under-prescriptive. For example, some believed they should be trusted to use their professional judgement and take a risk-based approach to issues. Otherwise, \textquote{GDPR is often like using a sledgehammer to crack nuts over things \textelp{} put barriers where they otherwise wouldn't need to be} (P9). Others wanted more precision about technical solutions and data retention periods. 
Despite their best efforts to be GDPR compliant, one marketer bemoaned, \textquote{how transparent is transparent? \textelp{} how much do you really have to spell it out \textelp{} to be really clear enough} (P8) after the Legal department had blocked their re-use of data collected during a campaign that had been designed to gather new leads.

A lawyer described the ambiguity that they experience whenever they suffer a data breach: \textquote{I regularly go to external counsel to get their view and they never have a definitive answer. It is always from experience, or we'll have to wait and see} (P7). Another lawyer described how they pore over ICO investigations to understand the decision-making and the findings that triggered the fines.

At the other end of the spectrum, people ask existential questions like what happens to GDPR in the UK after Brexit if there is a negative EU adequacy decision? One IT executive in a large company described it as \textquote{utterly bonkers} (P10) because \textquote{the damage it would do to both the UK and European economy would be just politically unacceptable.} The executive also thought they'd have to have two platforms--–UK and non-UK-–-if the EU failed to find the UK was offering an adequate level of data protection.

\subsubsection{The Data Audit Dividing Line}
\label{sec:challengeDataAudit}

Data audits distinguish the big from the small. When asked about the impact of data audits on their business, one SME responded, \textquote{What's a data audit} (P1). Two other SMEs were uncertain and assumed their GDPR IT outsourcer had taken care of it. On follow-up, one of the IT Services companies confirmed they stored the data and advised their clients, but \textquote{this is where it gets a little bit complicated \textelp{} they need to know where the PII is themselves} (P3).

Large companies approach it differently. They all do data audits. They find them time-consuming, but they appreciate they are \textquote{a good thing} (P7). One IT executive remarked, \textquote{It may be a pain in the backside, but once you've done it once, then at least you know where everything is. \textelp{} And you will be able to follow data around your organisation} (P12). Another lawyer described how they had undergone two audits–--in-house and external-–-and opined: \textquote{I found the audits helpful \textelp{} you can leverage off \textelp{} and show the reports to the directors and say either look how well I am doing in this area \textelp{} or we scored low here} (P7). Data audits are powerful tools in big business for building business cases for investment.

\subsubsection{The Weaponisation of Subject Access Requests (SARs)}
\label{sec:challengeSARs}

One SME has never received a SAR. Executives in another SME thought they hadn't, but it subsequently transpired that the CEO had handled a few personally. As companies scale up in size, the issue of SARs becomes more problematic. Disgruntled customers use SARs
\begin{displayquote}[P2]
as a stick to beat us with. They'll put in a SAR \textelp{} just to be awkward. They're saying \enquote{\textelp{} you have inconvenienced me, so now I'm going to inconvenience you.} Are they entitled to every internal email? They have rights to everything, but I'm saying, \enquote{but why? Why should they?} \textins{Perhaps} we'll do it offline \textins{in future}.
\end{displayquote}

Larger companies described similar issues with customers and, even more problematic, ex-employees. Some companies found it difficult to differentiate between emails that plainly referred to the ex-employee and deserved to be released and those that mentioned the ex-employee in a performance report alongside other employees. Other companies adopt a more proportionate response and require a precise aim. It is fair to say that companies are still finding their way.

\subsection{Suggested Improvements to GDPR}
\label{sec:suggestedImprovements}

At the end of the interviews, people were asked for their ideas on how could GDPR be made better. There was a certain amount of special pleading and wishful thinking. Nevertheless, the feedback points to ways in which GDPR could be made more accepted and more effective in achieving its goals.

\subsubsection{An SME-lite Version}
All the SMEs felt GDPR was overkill for companies like them that hold truly little data compared to Big Data companies. One CEO queried why they should be held to the same standard as a medical institution that holds sensitive personal data. \textquote{The rules I have to follow should not be the same ones as Goldman Sachs has to follow} (P5). The desire for simplification is understandable. Unfortunately the rights-based nature of GDPR hardly lends itself to differential watering down of protections for customers of SMEs but not of big business. Nevertheless, in practice, the regulator could consider applying the same principles on SMEs in a more proportionate manner.

\subsubsection{Reframe It}
A marketer suggested the regulator should demystify and reframe the message. \textquote{Less a pain in the arse type thing \textelp{} bring to the fore the real benefits \textelp{} in a more creative way} (P8). This may seem an unusual demand, but marketing spin is not alien to the EU Regulators. After all, most GDPR updates since 2019 usually include references to the benefits of the level playing field (LPF) and the competitive advantage to business of compliance. However, these messages do not resonate with this sample of companies. The LPF is irrelevant to SMEs who are typically domestic in focus and not material for larger companies if they already have operations in other countries. GDPR is not seen to confer a competitive advantage within the EU (because everyone must abide by it) and seen as potentially a disadvantage in non-EU countries if the competition is not saddled with the same restraints.

\subsubsection{Share It}
The GDPR expert in the bank felt the UK Regulator failed to support big business. \textquote{There is nothing to encourage people or companies to share best practice. There is not a forum \textelp{} or platform \textelp{} where the professionals can go and ask questions or share what works for them} (P11). At the other end of the expertise spectrum, the CEO of a SME felt let down for different reasons \textquote{I looked for checklists \textelp{} on government sites. Everyone is trying to get me to take a course to get a certificate in GDPR compliance} (P5). All they wanted to know was \textquote{what are the major things we should concentrate on} (P5).

\subsubsection{Clarify It}
Most respondents thought GDPR had brought legal clarity to the situation. Article 6 of GDPR is clear about the six lawful bases for one to process (collect, store, use etc.) people's data. However, the legal practitioners still felt there was a need for clarity on the wording in some instances, e.g., co-processor, international data transfers. \textquote{I did a Certificate in Data Protection Law \textelp{} and at one stage I was about as qualified as you could be, which was a bit of a joke, because I didn't know more than anybody else. You go to talk to a law firm \textins{about a case}. They have more experience but it's not necessarily they know more \textelp{} until there is more case law} (P7). Like the previous point about sharing learning, there seems a clear opportunity for the regulator to take a more proactive role in this area.

\subsection{Loosen it}
Some of the legal practitioners chafed against the rigidity of the rules. They argued that the regulator should allow a more commercial or risk-based approach of the rules for an informed professional such as happens with anti-money laundering. Questions remain about how this would work in practice including how such an approach is compatible with the fundamental rights character of data protection and how a risk-based approach could be made consistent. The other concern is the notion of risk itself and the risk thresholds (to the consumer?) that would need to be satisfied before GDPR could apply. 

\section{Discussion}
\label{sec:discussion}

\subsection{The benefits of GDPR to business}
Our findings show that GDPR has proven to be a windfall for companies who provide GDPR-related legal or technological advice and services. However, for the majority of companies where GDPR is not a core business, it still has had benefits. 

The threat of fines has changed the mindset of companies. In a world where data privacy is getting ever more important, GDPR has forced companies to catch up with their clients' desires and wishes to serve them only what they want to be served and use their data only in the way they want it to be used. It has forced companies to clean-up their act. This is good for companies and society. 

The threat of fines has changed the data infrastructure of companies. In a world where compliance projects trump non-compliance projects, GDPR has forced companies to modernise and upgrade their data management, data quality and information security. GDPR has gifted companies a reason to invest in projects, such as rationalising legacy databases, that they knew were important but kept putting on the long finger. It has delivered many of the `usual' benefits of an IT project directly to companies whose technology was sub-optimal and it has indirectly benefited companies whose technology was adequate but still required enhancements to meet the regulations. 

GDPR has delivered some standardisation benefits, and it has been used by some companies as a means to signal their privacy credentials to the benefit of their brand and reputation. 

Our findings on benefits do not tally with many of the projected benefits in earlier literature. The area of agreement is around improved data management process~\cite{bennett_european_2018,fimin_council_2018}, use of analytics~\cite{garber_gdpr_2018} and increased security~\cite{krikke_gdpr_2019}. 
There is some equivocal overlap in the area of reputational enhancement~\cite{beckett_gdpr_2017,tikkinen-piri_eu_2018}. We found some marketing participants shared the same belief. However, many of the other assertions such as improved consumer confidence~\cite{dellie_gdpr_2019} and trust ~\cite{dellie_gdpr_2019,dubrova_challenges_2018}, legal clarification~\cite{dubrova_challenges_2018}, competitive advantage~\cite{dellie_gdpr_2019} and cost reduction~\cite{perry_gdpr_2019,miglicco_gdpr_2018,beckett_gdpr_2017,obrien_privacy_2016} were not supported by our findings. The size of the GDPR fining system was well understood in advance but none of the literature seems to have picked up the transformational effect it was going to have on corporate psychology.

\subsection{There are winners and losers}
Our findings show that the benefits of GDPR have fallen differently within a business. It has created new power bases within companies. Depending on the industry, it will have a different name, but typically GDPR expertise sits in the Risk, Compliance or Legal department and the IT/IS or Information Security department. Both have enjoyed boosts to budgets and headcount. Suffering a high-profile data breach that could destroy a company's reputation and potentially suffering a big-ticket fine that could ruin a company's finances has meant that GDPR risk continues to be a board agenda item. This means both departments continue to be more involved with corporate-level data decision-making than before. In addition, the Marketing department has a higher quality, more up-to-date database of customers and their communication preferences. 

So, while Legal and IT may be winners, are there losers? Yes. There are direct and indirect losers. The most obvious are the executives spread across an organisation who championed projects that were delayed or killed in competition with higher priority GDPR initiatives. Less obvious are senior management. Their discretion was hemmed in pre-GDPR by the need to prioritise GDPR-readiness. Their discretion is now policed by Legal or IT departments who follow breach investigations zealously and remind them that GDPR compliance is an ongoing commitment. The indirect losers are the departments that have to handle the extra workload generated by GDPR compliance, e.g., Human Resources having to negotiate with disgruntled ex-employee SARs, Customer Service having to deal with dissatisfied customer SARs and Marketing having to constantly update customer notification preferences. A lawyer said \textquote{I think if you were to ask a marketing person what are the benefits \textelp{} I think they might struggle to articulate some benefits} (P7). Another lawyer characterised the perception of their role and GDPR, \textquote{From a marketing perspective \textelp{} they see it as a stopper} (P11). 

A lasting legacy of GDPR is a shift of power. It has put non-commercial functions, who were hitherto regarded as support functions, in the heart of strategic commercial decision making. The long-term implications of this remain to be seen.

To the best of our knowledge, this change in power-dynamics has not been anticipated in earlier literature.

\subsection{Implementation issues remain}
GDPR is not without its dis-benefits. This was not the primary focus of our research but we identify a number of challenges. New business development and intra-company communication is more constrained. Regulation has increased costs and internal bureaucracy. Grey areas remain due to a lack of case law. Disgruntled customers and employees weaponise SARs as a tool of retaliation.

Our findings on challenges tally with many of the issues identified in earlier literature. The complexity of GDPR, its lack of specificity, its subjectivity, the cost overhead, the difficulty recruiting and retaining expert staff and operationalising the right to erasure were all well anticipated. The restrictions on marketing were known in theory but the effect on new business development in practice was underappreciated. The chilling effect on intra-company communication does not appear in earlier literature. On the other hand, some hypothesised downsides, such companies cutting back their service offerings in the EU to avoid GDPR, did not ever come up in conversation. 

When we asked our participants for their own ideas as to how to improve GDPR, we find that they believe that regulators should re-frame GDPR messaging to be more positive, sponsor forums to facilitate the sharing of learning and coping strategies, clarify policies and apply lighter standards on small business. 

Existing literature does not consider getting business buy-in to GDPR. The emphasis is more on the punitive power of GDPR. In contrast, the literature has long recognised the need to simplify and clarify its requirements. Clamour for a more SME-friendly version however is relatively new.

\subsection{ePrivacy Regulation}
There is another EU privacy regulation that will impact business: the ePrivacy Regulation (ePR)~\cite{europeandataprotectionboard_statement_2021}. Commonly referred to as the `cookie law', it was due to take effect in parallel with the GDPR in 2018, but it has been repeatedly delayed and redrafted due to fierce lobbying and institutional debate. The 14th draft was approved by the Council in early 2021 and passed to the European Parliament for review. If approved, a two-year transitional period would apply to allow organisations the time to adjust to the new requirements. As such, the earliest it could come into force is late 2023.

The ePR is designed to complement GDPR and fit in the same framework. It has a narrower focus and applies to privacy in electronic communications. ePR is a \textit{lex specialis} to GDPR which means it will override GDPR if there is ever a contradiction.

The ePR applies to new players such as WhatsApp, Facebook and Skype to guarantee the same level of confidentiality as traditional telecommunications operators. Privacy is guaranteed for communications content and, significantly, metadata. There will be simpler rules on cookies to address the consent fatigue that has become the bane of GDPR. The new rules will require browsers to provide an easy way to accept or refuse tracking cookies and identifiers. It will also clarify that no consent will be needed for non-intrusive cookies that improve the internet experience, such as cookies that remember one’s shopping cart history. In addition, it proposes to ban spam. And it will carry the same fining system and regulation status as GDPR.

The ultimate impact on companies such as those in our survey is hard to predict. The only certainty is that it will generate more advisory business for the law firm and IT suppliers. In theory, it may mean fewer cookie consent pop-ups, but it may also mean more pop-ups saying \textquote{sorry, no entry if no cookies} or even \textquote{sorry, pay a premium if no cookies}. Thus we can imagine our bank being relaxed about fewer cookies (they know their clients), the publisher being more hard-line about cookie capture (they need the data brokerage revenue) and the remainder torn between wanting to make the online experience free of pop-ups but wanting consent to capture the metadata and track consumer behaviour. The final shape of the ePR is far from agreed. For example, the direct marketing industry is still negotiating regarding cookie paywalls, soft opt-ins, and whether it will apply to B2B marketing since GDPR protects the personal data of `natural persons' but not the personal data of `legal persons' such as a company~\cite{bateman_eprivacy_2021}.

\section{Conclusion}
\label{sec:conclusion}

GDPR is a regulation that is designed to safeguard EU citizens' data privacy. The benefits to the consumer and the regulator and the downsides to business are relatively predictable. What we are interested in however is whether there are any benefits of GDPR to business and how they might affect the different parts of an organisation. To our knowledge, nobody has looked at this from the perspective of business since GDPR came into effect in May 2018. 

Using semi-structured interviews, we surveyed 14 C-level executives responsible for business, finance, marketing, legal or IT drawn from 6 small, medium and large companies in the UK and Ireland. We deliberately sampled beyond the IT department, which tends to be the typical target of GDPR surveys, to obtain a fuller picture.  

We find the threat of large fines has focussed the minds of business and made it more privacy conscious. GDPR has gifted companies a reason to justify investment in modernising their data management processes and security. Companies have cleaner and more up-to-date customer databases. In the absence of GDPR, companies admit they would ask for more information than necessary, use it more frequently, hold it for longer and keep it less securely.

It has created new power bases within organisations that act as guardians or champions of privacy. Such in-house regulators will continue to enjoy influence on corporate decision making provided the official regulators maintain a steady news flow on enforcement actions against offenders and data breaches.

We find that many implementation issues exist that would benefit from better communication, guidance and simplification by the EU and its regulatory arm. 

In summary, GDPR may be a headache to business but it has made it more careful with data. Judged by that standard, GDPR has been a successful socio-technical regulation. 

\subsection{Limitations}
There was little empirical research on this topic with which to compare and contrast our findings. The number of companies and participants in our survey is small. It suffers from a sampling bias despite our best attempts to get a balanced selection from across a company's entire senior management team. Alas, ask a company to take part in a GDPR study and they will invariably point to their GDPR experts. 

\subsection{Future Work}
It would be a natural next step to expand and extend the study to other sectors and geographies to see if the themes identified in this study also apply. It will be interesting to observe how the power dynamics may evolve within companies as the real risk of fines becomes evident over time. Open questions surround how the application of GDPR could be made more proportionate in its application to SMEs whilst maintaining consistent protection of consumers rights.

\printbibliography
\balance

\clearpage

\appendix
\section{Interview framework} 

\begin{itemize}
    \item \textbf{Respondent: Tell me about your job}
    \begin{itemize}
        \item   Industry sector \& Size of company?
        \item 	What is your role, your title, your department?
        \item 	What does your company/department use customer data for? Describe 
    \end{itemize}    
    \item \textbf{Open questions: Do you know GDPR?}
     \begin{itemize}   
        \item How has GDPR affected your day-to-day work/department/division/company? 
        \item 	Grey areas?
        \item 	Biggest benefits?
        \item 	Biggest challenges?
    \end{itemize}
    \item \textbf{Benefits checklist: Tell me about the good things}
     \begin{itemize}   
        \item 	Improved your company's brand/reputation?
        \item 	Is customer trust higher? How do you know?
        \item 	Is there more legal certainty? Grey areas?
        \item 	Level playing field across Europe; Has it translated into a bigger market for your company?
        \item 	Has GDPR compliance conferred a competitive advantage - in EU and in non-EU markets?
        \item 	Has it spurred innovation - from data portability and data flows, new products/services?
        \item 	Has GDPR led to growth? Incentives to invest more?
        \item 	Better/different advertising?
        \item 	Did it result in an upgrade to internal systems/streamlined processes?
        \item 	Is security better?
    \end{itemize}
        
        \item \textbf{Challenges: Tell me about the bad things}
         \begin{itemize}
        \item 	Departmental drawbacks/limitations?
        \item 	Negative company-wide and/or market impacts?
        \item 	Compliance cost overheads?
        \item 	Data audit impacts?
        \item 	Data minimisation impacts?
        \item 	Data security rigidity?
        \item 	Data breaches/Greater liabilities to fines?
        \item Accountability and governance: how does it work? Bureaucratic?
        \item 	Privacy rights: satisfying SARs, right to correction and deletion?
        \item 	Are privacy-first processes making it harder to advertise/market/service customers?
          \end{itemize}

        \item\textbf{Improvements}
        \begin{itemize}
        \item 	What makes good GDPR good?
        \item 	Any recommendation?
         \end{itemize}
        \item \textbf{What would you do different today if there was no GDPR?}
\end{itemize}

\balance

\end{document}